\documentclass[twoside]{dis07}
\usepackage[latin1]{inputenc}
\usepackage[dvips]{graphicx,epsfig,color}
\usepackage{wrapfig,rotating}
\usepackage{amssymb,amsmath,array}

\begin{document}
\title{Impact of and constraints on PDFs at the LHC}

\author{A M Cooper-Sarkar 
%
%
\vspace{.3cm}\\
%
University of Oxford - Dept of Physics \\
Denys WIlkinson Building, Keble Rd, Oxford, OX1 3RH - UK
%
}

\maketitle

\begin{abstract}
Uncertainties on parton distribution functions (PDFs) compromise discovery at the LHC for 
any new physics which can be described as a contact-interaction.
PDF uncertainties also limit our ability to use $W$ and $Z$ cross-sections
as anaccurate luminosity monitor. The impact of the current level of PDF 
uncertainty on LHC physics is reviewed and the possibility of reducing this 
uncertainty using LHC data is investigated. 
\end{abstract}

\section{Impact of PDF uncertainties on $W$ and $Z$ production}
\label{sec:intro}

\begin{figure}[tbp]
\vspace{-2.0cm} 
\centerline{
\epsfig{figure=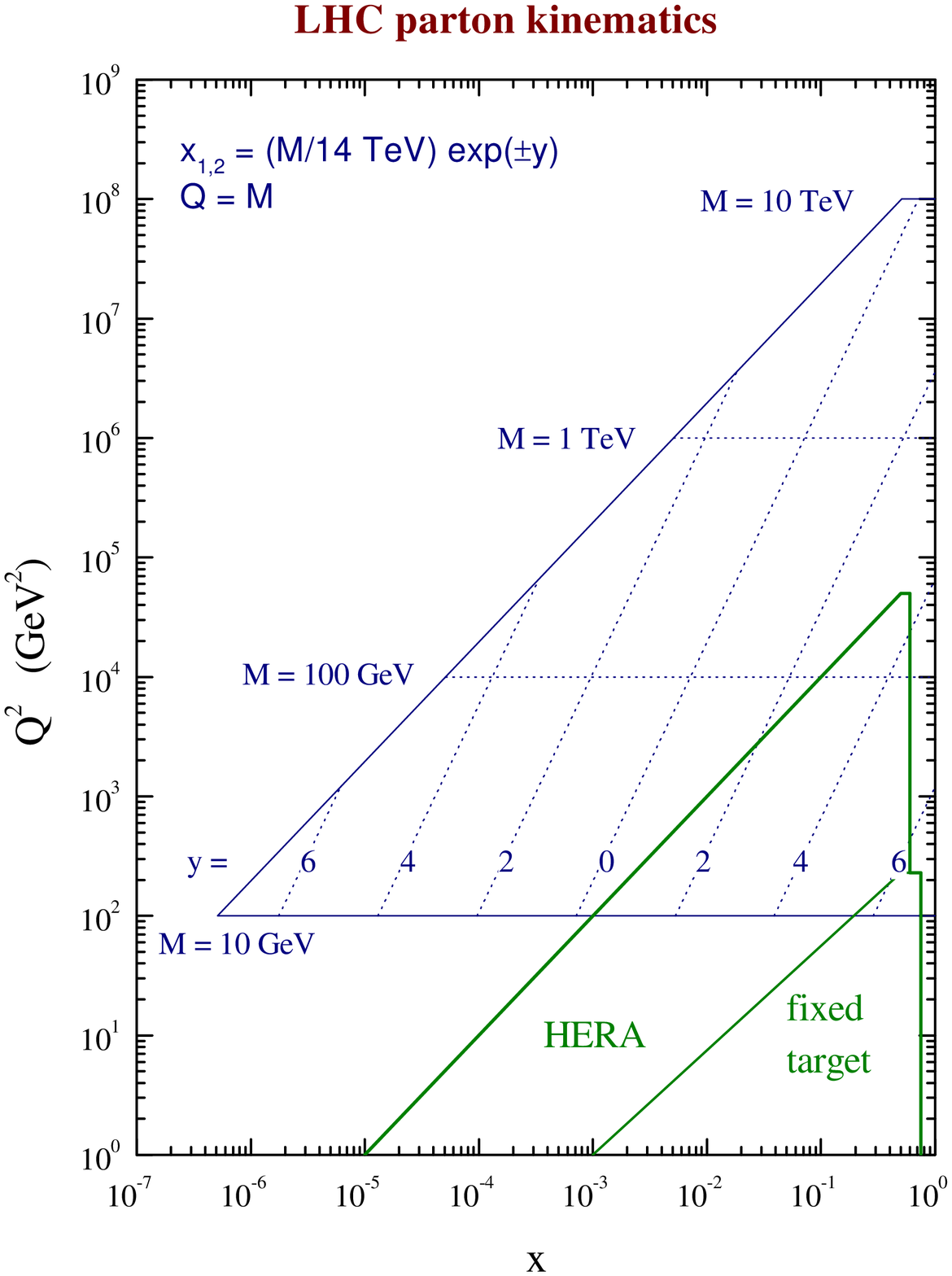,width=0.5\textwidth}
\epsfig{figure=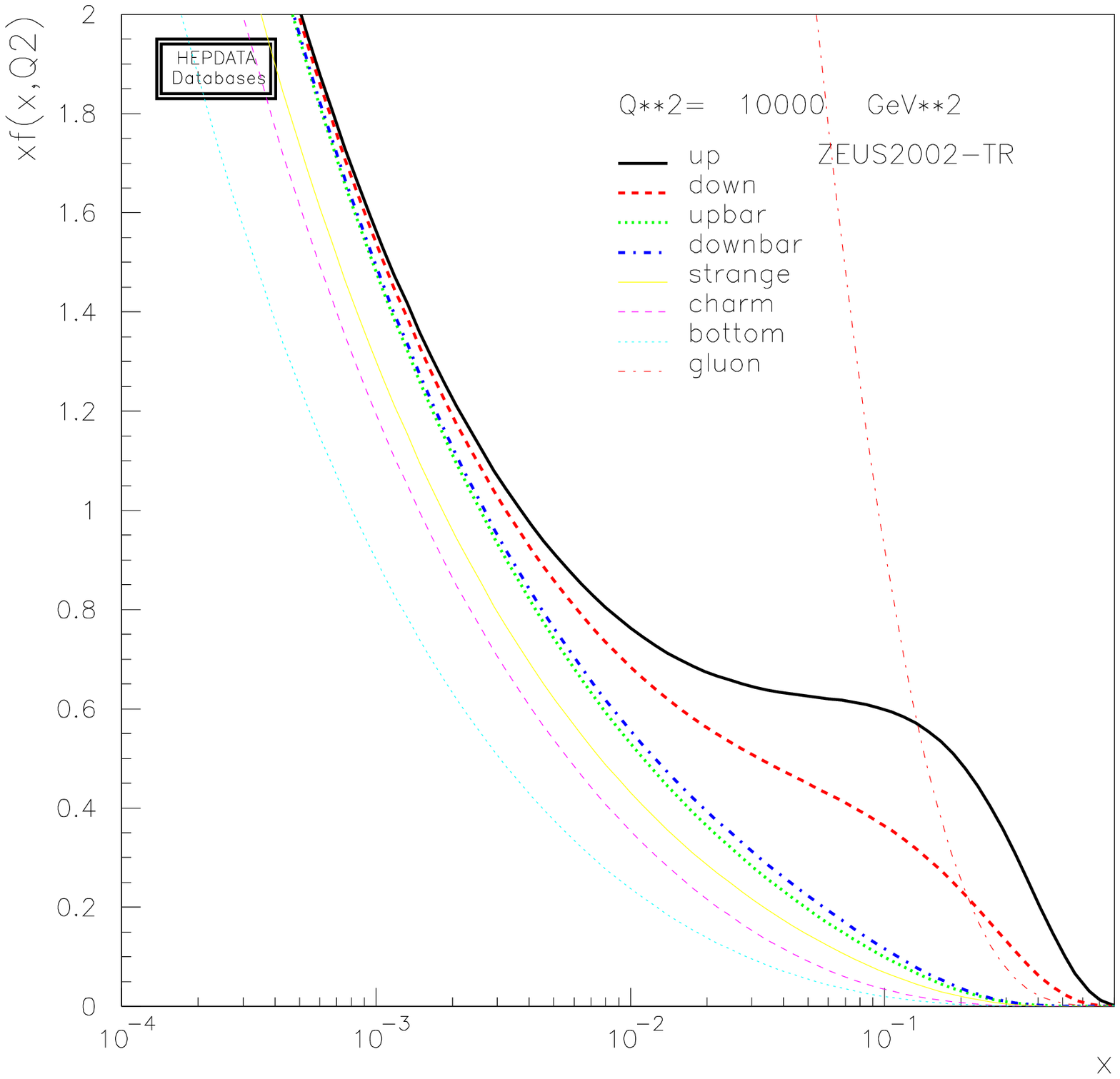,width=0.5\textwidth}}
\caption {Left plot: The LHC kinematic plane (thanks to J Stirling).
Right plot: PDF distributions at $Q^2 = 10,000$~GeV$^2$.}
\label{fig:kin/pdfs}
\end{figure}

The kinematic plane for LHC parton kinematics is shown in 
Fig.~\ref{fig:kin/pdfs}. At leading order (LO), $W$ and $Z$ production occur 
by the process, $q \bar{q} \rightarrow W/Z$, and the momentum fractions of the 
partons participating in this subprocess are given by, 
$x_{1,2} = \frac{M}{\surd{s}} exp (\pm y)$, where $M$ is the centre of mass 
energy of the subprocess, $M = M_W$ or $M_Z$, $\surd{s}$ is the centre of
 mass energy of the reaction  ($=14$ TeV at the LHC) and 
$y = \frac{1}{2} ln{\frac{(E+pl)}{(E-pl)}}$ gives the parton rapidity. 
Thus, at central rapidity, the participating partons have small momentum 
fractions, $x \sim 0.005$.
Moving away from central rapidity sends one parton to lower $x$ and one 
to higher $x$, but over the measurable rapidity range, $|y| < 2.5$, $x$ values 
remain in the range, $5\times 10^{-4} < x < 5\times 10^{-2}$. Thus the 
scattering is happening between sea quarks. Furthermore, the high 
scale of the process $Q^2 = M^2 \sim 10,000$~GeV$^2$ ensures that the gluon is 
the dominant parton, see Fig.~\ref{fig:kin/pdfs}, so that these sea quarks 
have mostly been generated by the flavour blind $g \to q \bar{q}$ splitting 
process. Thus the precision of our knowledge of $W$ and $Z$ cross-sections at 
the LHC is crucially dependent on the uncertainty on 
the momentum distribution of the gluon. 

The cross-sections for  $W/Z$ production have 
been suggested as `standard-candle' processes for luminosity measurement, 
because theoretical uncertainties are well controlled, and the uncertainty from the PDFs was
thought to be small  (e.g. MRST01 PDFs\cite{mrst} predict an impressive $2\%$ uncertainty).
However, when considering the PDF uncertainties it is necessary not 
only to consider the uncertainties of a particular PDF analysis but also to compare PDF 
analyses. Predictions for the $W/Z$ cross-sections, decaying to the lepton decay mode, 
are given for modern PDF sets in Table~\ref{tab:datsum}.
\begin{table}[t]
\centerline{\small
\begin{tabular}{llllcccc}\\
 \hline
PDF Set  & $\sigma(W^+).B(W^+ \rightarrow l^+\nu_l)$ & $\sigma(W^-).B(W^- \rightarrow l^-\bar{\nu}_l)$ & 
$\sigma(Z).B(Z \rightarrow l^+ l^-)$\\
 \hline
 ZEUS-JETS~\cite{zeujets}  & $11.87 \pm 0.45 $~nb & $8.74 \pm 0.34 $~nb & $1.89 \pm 0.07$~nb \\
 ZEUS-S~\cite{zeus-s}  & $12.07 \pm 0.41 $~nb & $8.76 \pm 0.30 $~nb & $1.89 \pm 0.06$~nb\\
 CTEQ6.1~\cite{cteq} & $11.66 \pm 0.56 $~nb & $8.58 \pm 0.43 $~nb & $1.92 \pm 0.08$~nb\\
CTEQ6.5~\cite{cteq65} & $12.44 \pm 0.47 $~nb & $9.12 \pm 0.36 $~nb & $2.04 \pm 0.07$~nb\\
 MRST01~\cite{mrst} & $11.72 \pm 0.23 $~nb & $8.72 \pm 0.16 $~nb & $1.96 \pm 0.03$~nb\\
 MRST04~\cite{mrst04} & $11.74 \pm 0.23 $~nb & $8.71 \pm 0.16 $~nb & $1.96 \pm 0.03$~nb\\
 \hline\\
\end{tabular}}
\caption{LHC $W/Z$ cross-sections for decay via the lepton mode, 
for various PDFs}
\label{tab:datsum}
\end{table}

Whereas the $Z$ rapidity distribution can be fully reconstructed from 
its decay leptons, this is not possible for the $W$ rapidity distribution, 
because the leptonic decay channels which we use to identify the $W$'s have 
missing neutrinos. Thus we actually measure the $W$'s 
decay lepton rapidity spectra rather than the $W$ rapidity spectra.
\begin{figure}[tbp] 
\vspace{-3.0cm}
\centerline{
\epsfig{figure=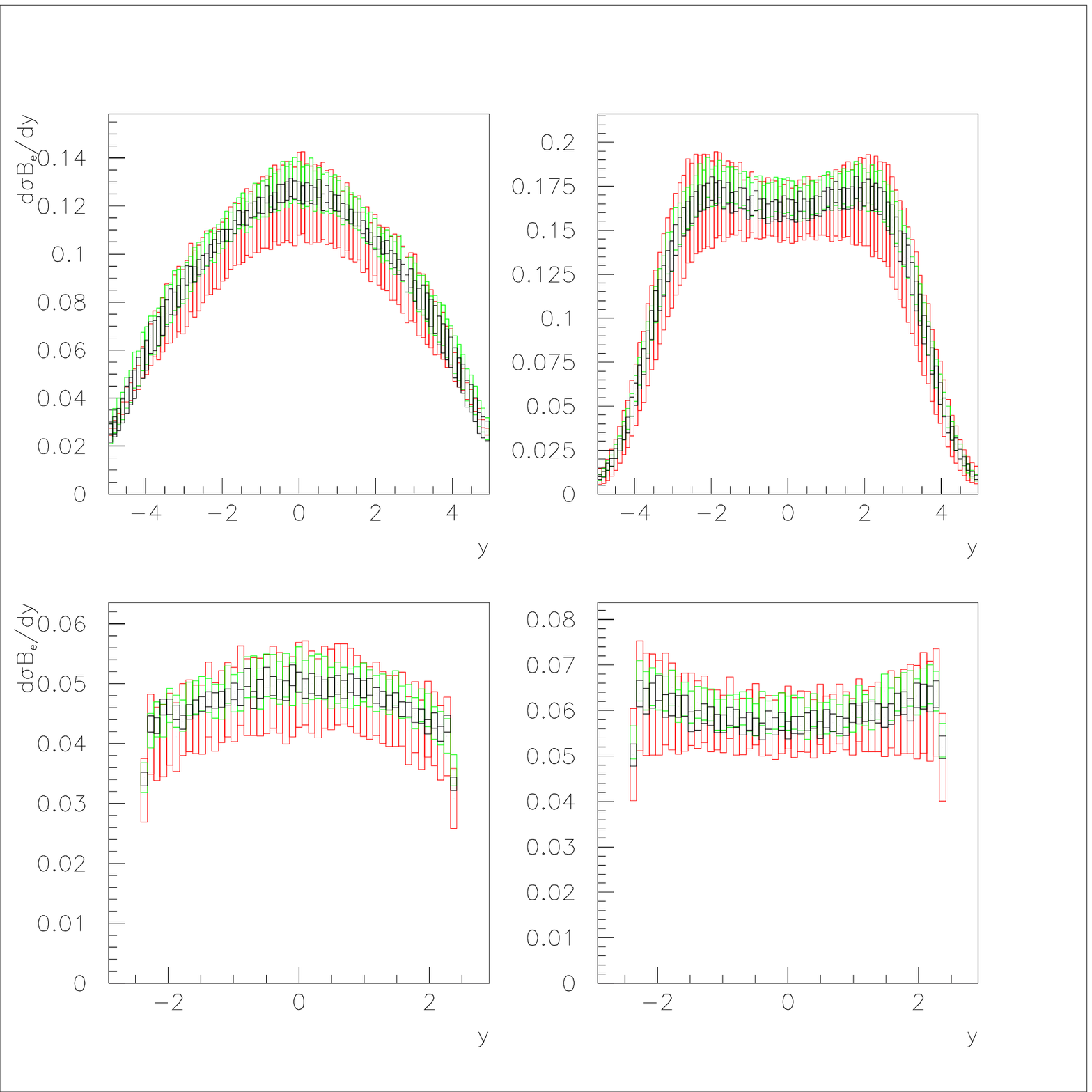,width=0.66\textwidth,height=7cm }
\epsfig{figure=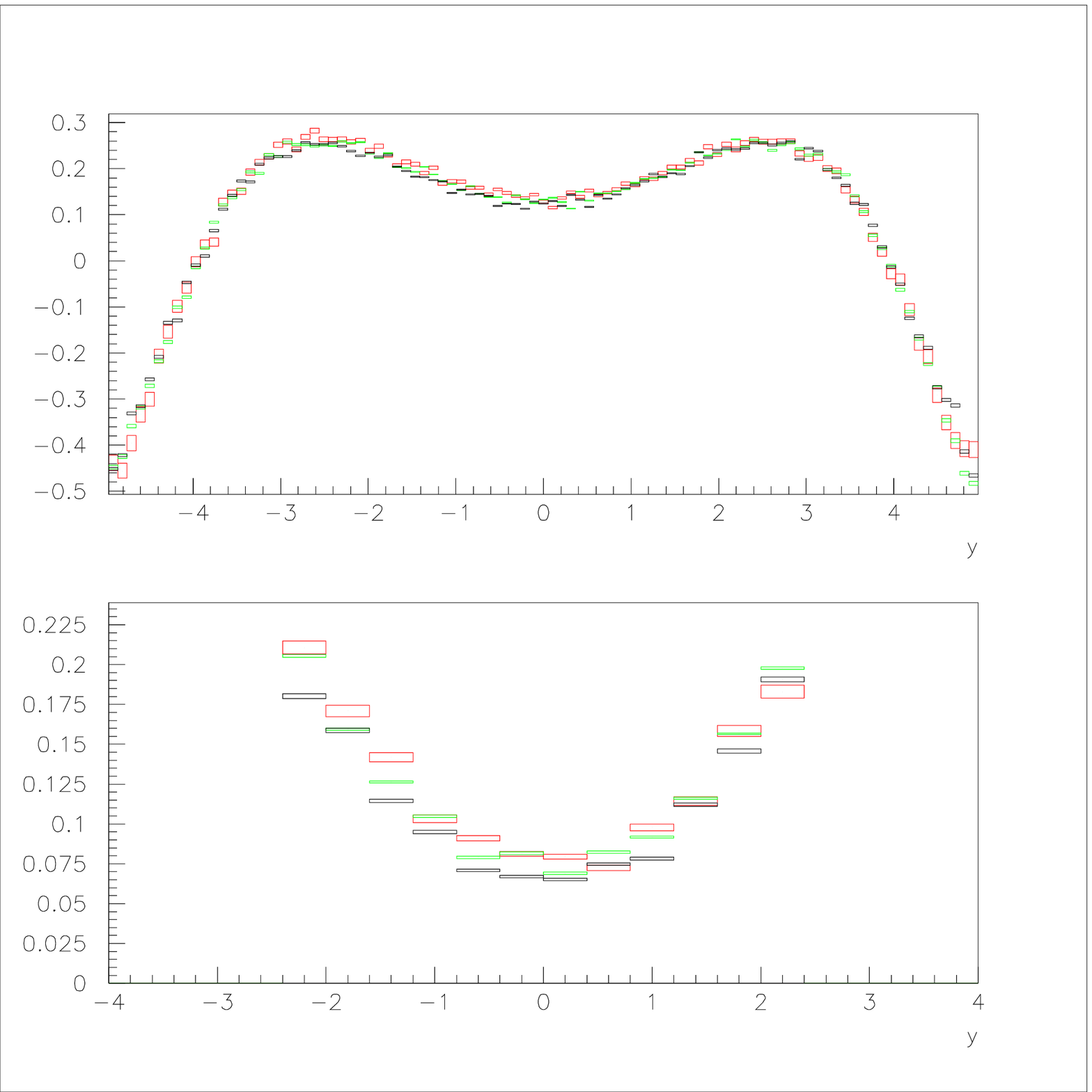,width=0.33\textwidth,height=7cm}
}
\caption {Top row: $e^-$, $e^+$ and $A_e$ rapidity spectra for the lepton from the $W$ decay, 
generated using HERWIG + k factors and CTEQ6.1 (red),
ZEUS-S (green) and MRST2001 (black) PDF sets with full uncertainties. Bottom row: the same spectra after passing 
through the ATLFAST~\cite{atlfast} detector simulation and selection cuts.(Thanks to A Tricoli)}
\label{fig:gendet}
\end{figure} 
Fig.~\ref{fig:gendet} compares the predictions for the lepton spectra from $W^{\pm}$ decay
for the ZEUS-S~\cite{zeus-s} PDFs with those of 
the CTEQ6.1\cite{cteq} PDFs and the MRST01 
PDFs\footnote{MRST01 PDFs are used because the 
full error analysis is available only for this PDF set.}. 
This figure is based on one million simulated, $W \rightarrow e \nu_e$, events for each 
of the PDF sets using HERWIG (6.505)~\cite{herwig}. 
For each of these PDF sets the eigenvector error PDF sets~\cite{LHAPDF} 
have been simulated 
by PDF reweighting and k-factors have been applied to approximate an NLO generation~\cite{tricoli}.
The top part of Fig.~\ref{fig:gendet}, shows the $e^{\pm}$ spectra at the generator level. 
The events were then passed through the ATLFAST fast simulation of the ATLAS detector, 
which smears the 4-momenta of the leptons to mimic momentum dependent detector resolution. 
The following 
selection cuts are then applied: pseudorapidity, $|\eta| <2.4$, to avoid bias at the edge of 
the measurable rapidity range; $p_{te} > 25$ GeV, since high $p_t$ is necessary for electron 
triggering; missing $E_t > 25$ GeV, since the $\nu_e$ in a signal event will have a 
correspondingly 
large missing $E_t$; no reconstructed jets in the event with $p_t > 30$ GeV and recoil on the 
transverse plane $p_t^{recoil} < 20$ GeV, to discriminate against QCD background.
The lower half of Fig.~\ref{fig:gendet}, shows the $e^{\pm}$ spectra at the detector level 
after application of cuts and smearing. Comparing the uncertainty at central rapidity, rather 
than the total cross-section, we see that the uncertainty estimates are 
rather larger: $~5\%$ for ZEUS-S; $~8\%$ 
for CTEQ6.1M and about $~3\%$ for MRST01. Considering both 
Fig.~\ref{fig:gendet} and Table~\ref{tab:datsum} we conclude that
the spread in the predictions of these different PDF sets is 
larger than the uncertainty estimated by the individual analyses. 
Currently the overall uncertainty of these NLO predictions is $\sim 8\%$
This suggests that measurements which are accurate to $\sim 4\%$ 
could discriminate between PDF sets.

\begin{figure}[tbp] 
\centerline{
\epsfig{figure=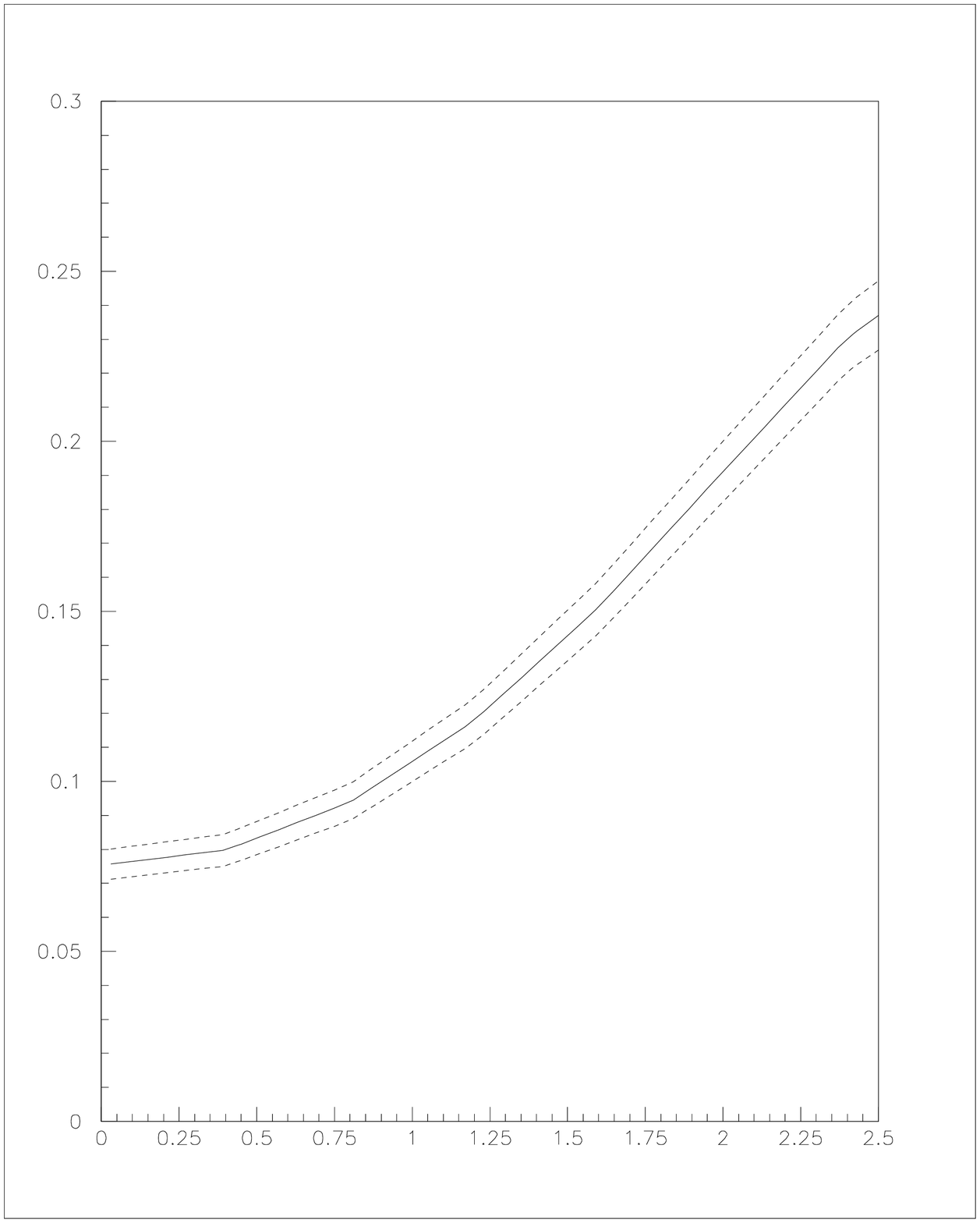,width=0.33\textwidth,height=3.5cm }
\epsfig{figure=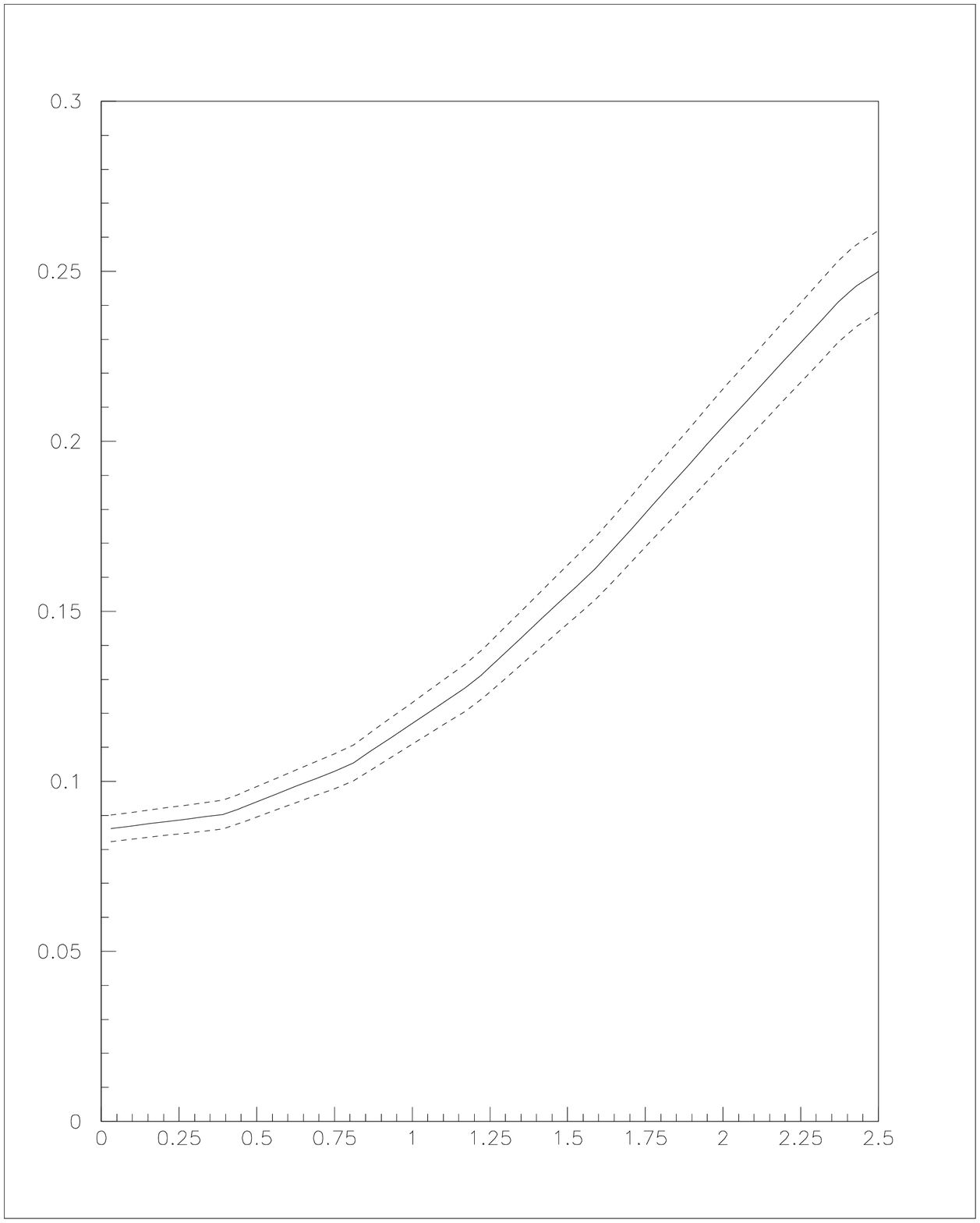,width=0.33\textwidth,height=3.5cm}
\epsfig{figure=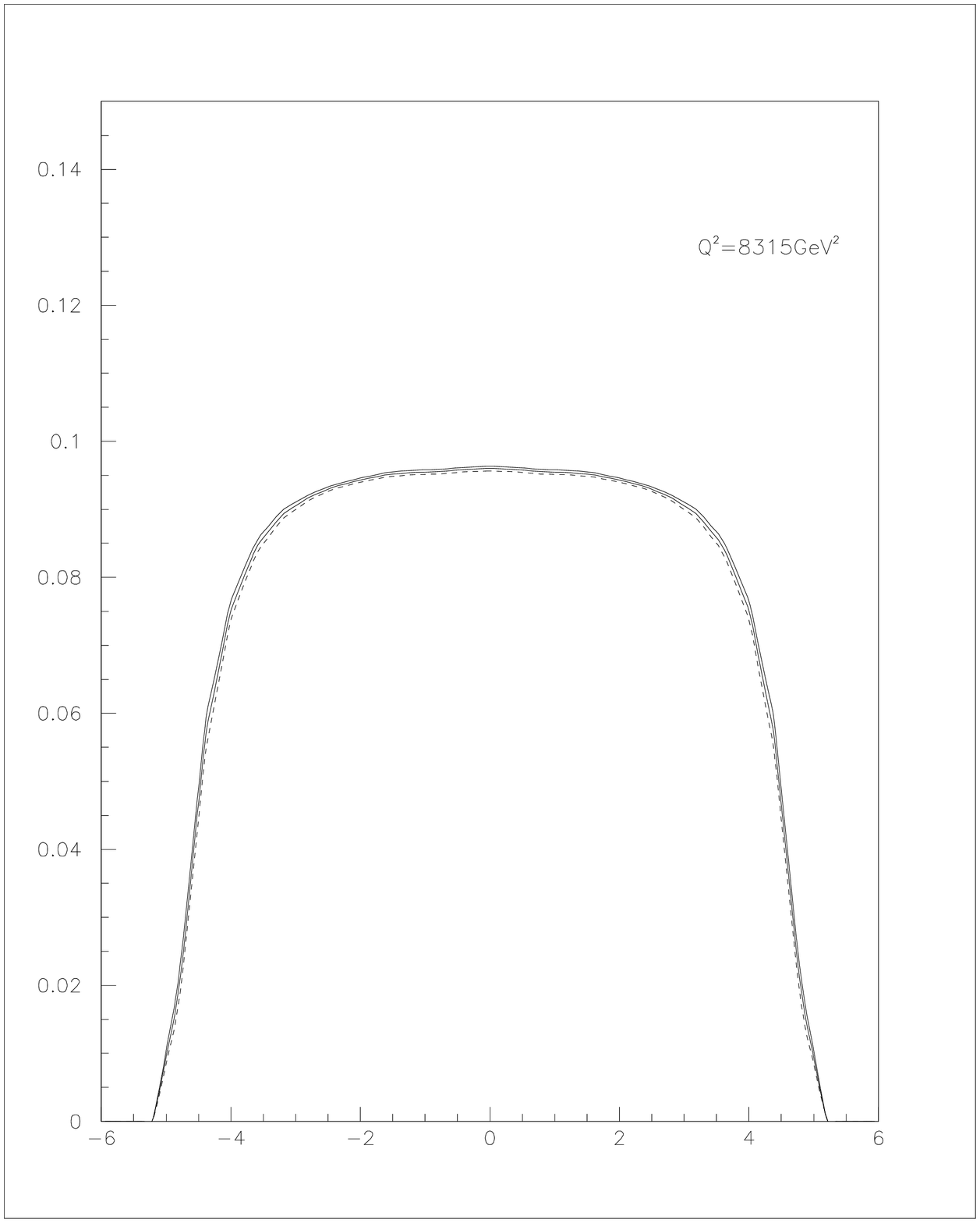,width=0.33\textwidth,height=3.5cm}
}
\caption {The lepton asymmetry, $A_e$, as predicted by MRST04 (left) and CTEQ6.1 (centre) PDFs.
 Right: the $Z/W$ ratio, $A_{ZW}$, as predicted by CTEQ6.1 PDFs}
\label{fig:awzwlepton}
\end{figure}
Since the PDF uncertainty feeding into the $W^+, W^-$ and $Z$ production is 
dominated by the gluon PDF, for all three processes, there is a 
strong correlation in their uncertainties, which can be 
removed by taking ratios.  The PDF uncertainties on the $W$ asymmetry 
\[A_W = (W^+ - W^-)/(W^+ + W^-).\] at central rapidity are dependent on the $u$ and $d$ 
valence PDFs at small $x$, $x \sim 0.005$. This is simply understood from LO QCD 
\[A_W = (u_v - d_v)/(u_v +d_v+ \bar{u} + \bar{d}).\] We will actually measure the lepton 
asymmetry, \[A_l = (l^+ - l^-)/(l^+ + l^-).\] and predictions for $A_l$ from MRST04 and CTEQ6.1 
PDFs are shown in Fig.~\ref{fig:awzwlepton}.  A difference of $\sim 13\%$ at central rapidity 
orginates in a difference of the their valence parametrizations at low-$x$~\cite{url1,url2} 
Fig.~\ref{fig:gendet} also
shows predictions for the  lepton asymmetry, at generator and at detector level.
A particular lepton rapidity can be fed from a range 
of $W$ rapidities so that the contributions of partons at different $x$ values 
is smeared out in the lepton spectra. The cancellation of the 
uncertainties due to the gluon PDF is not so perfect in the lepton asymmetry 
as in the $W$ asymmetry, nevertheless the sensitivity to $u$ and $d$ quark 
valence distributions remains. Hence LHC measurements of the lepton asymmetry at central 
rapdity should give information on the valence distributions at small-$x$, $x \sim 0.005$, 
where there are currently no measurements.

PDF sensitivity can be removed almost completely by taking the ratio \[A_{ZW} = Z/(W^+ +W^-)\] as
shown on the right hand side of Fig.~\ref{fig:awzwlepton}. The figure has been made using 
CTEQ6.1 PDFs, but all modern PDF sets give an indistinguishable result.
The PDF uncertainties on this quantity are as small as $\sim 1\%$. 
Hence this ratio can be used as a 
benchmarks for our understanding of Standard Model Physics at the LHC.

\section{Using LHC data to improve precision on PDFs}

\begin{figure}[tbp] 
\vspace{-1.0cm}
\centerline{
\epsfig{figure=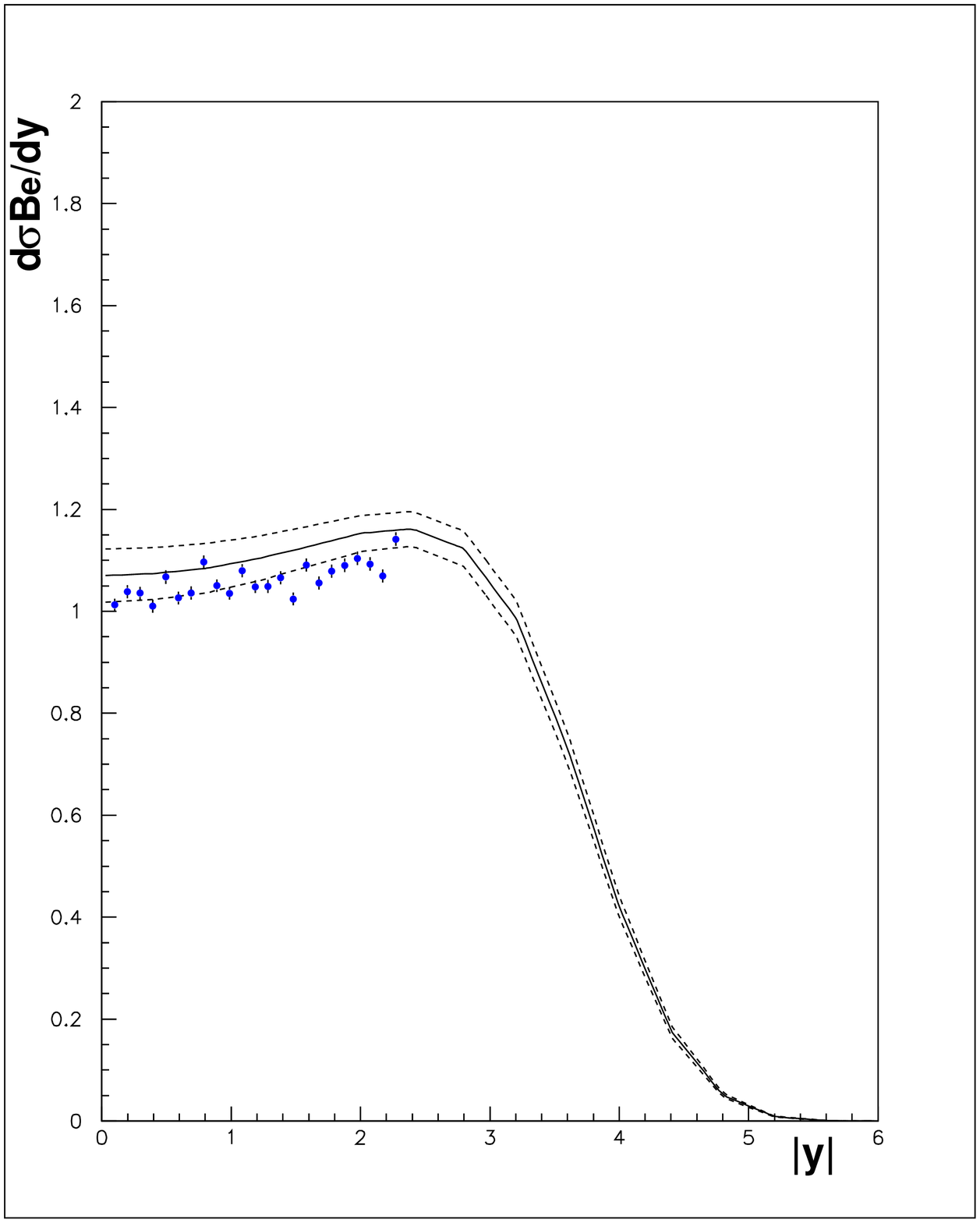,width=0.3\textwidth,height=4cm}
\epsfig{figure=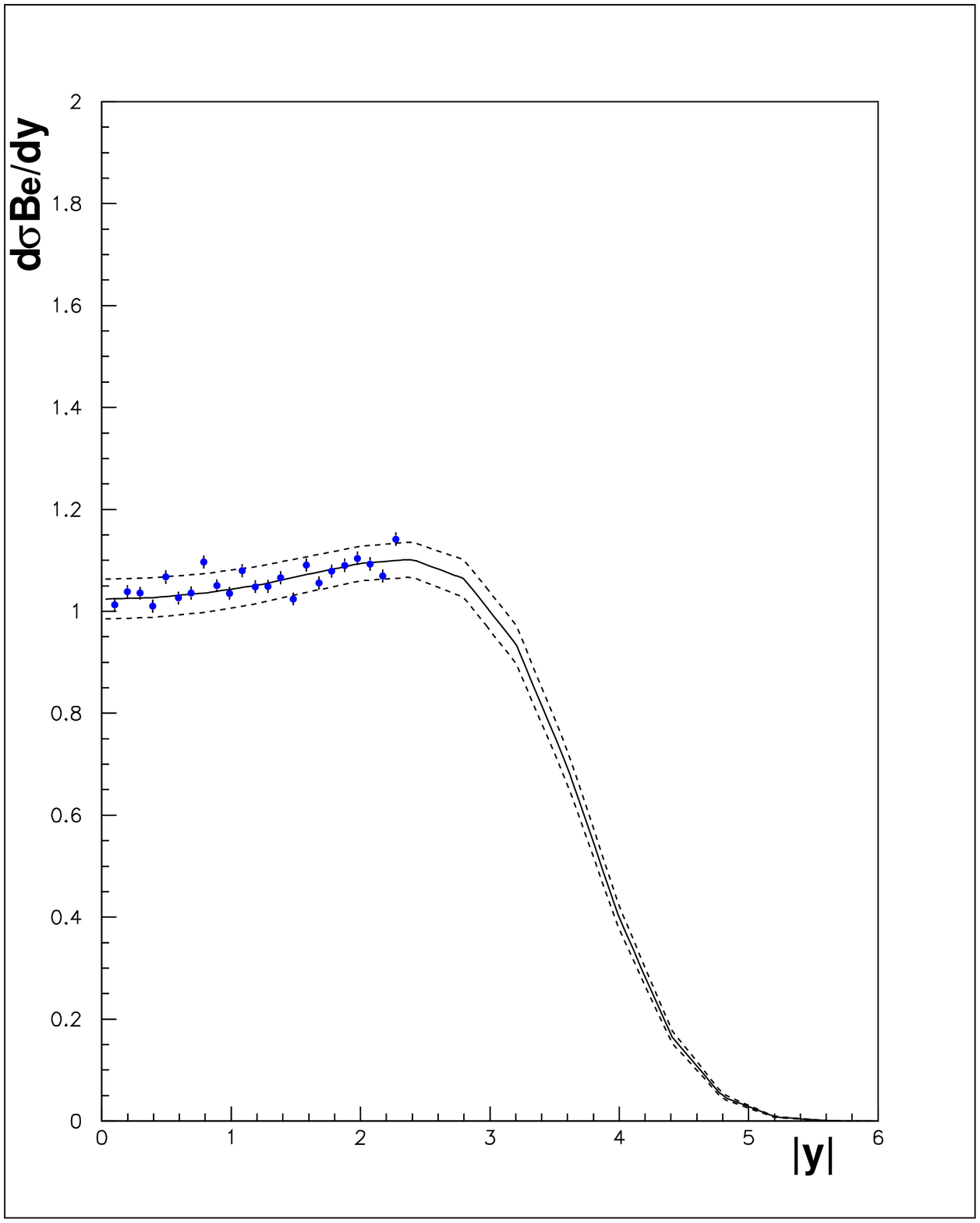,width=0.3\textwidth,height=4cm}
}

\caption {Left: $e^+$ rapidity spectrum generated from CTEQ6.1 PDFs, which have been passed through 
the ATLFAST detector simulation and corrected back to generator level using ZEUS-S PDFs, 
compared to the analytic prediction
using ZEUS-S PDFs. Right: the same lepton rapidity spectrum compared to the analytic 
prediction AFTER including these lepton pseudo-data in the ZEUS-S PDF fit.}
\label{fig:ctqcorfit}
\end{figure}

 The high cross-sections for $W$ production at 
the LHC ensure that it will be the experimental systematic errors, 
rather than the statistical errors, which 
are determining. We have imposed a random  $4\%$ scatter on our samples of 
one million $W$ events, generated using different PDFs, in order to 
investigate if measurements at this level of precision 
will improve PDF uncertainties at central rapidity significantly if they 
are input to a global PDF fit. Fig.~\ref{fig:ctqcorfit} shows the $e^+$ 
rapidity spectra for events generated from the CTEQ6.1 PDFs 
($|\eta| < 2.4$) and passed through the ATLFAST detector simulation and 
cuts. These data are then corrected back from detector level 
to generator level using a different PDF set- in this cases the ZEUS-S PDFs- 
since in practice we will not know 
the true PDFs. 
On the left hand side of the figure these data are compared to the analytic 
predictions from the ZEUS-S PDFs. The right hand side of the figure 
illustrates the result if these pseudo-data are then 
included in the ZEUS-S PDF fit. The central value of the fit prediction 
shifts, showing its sensitivity to the new data and the size of the PDF 
uncertainties at central rapidity decreases from $6\%$ to $4.5\%$.  
The largest shift and improvement in uncertainty is in the PDF parameter 
$\lambda_g$ controlling the low-$x$ gluon at the input scale, $Q^2_0$: 
$xg(x) \sim x^{\lambda_g}$ at low-$x$, $\lambda_g = -0.199 \pm 0.046$, 
before the input of the LHC pseudo-data, compared to, 
$\lambda_g = -0.181 \pm 0.030$, after input. 
Note that whereas the relative normalisations of the $e^+$ and $e^-$ spectra 
are predicted by the PDFs, the absolute normalisation of the data is free in the 
fit so that no assumptions are made on our ability to measure luminosity.

\begin{figure}[tbp] 
\centerline{
\epsfig{figure=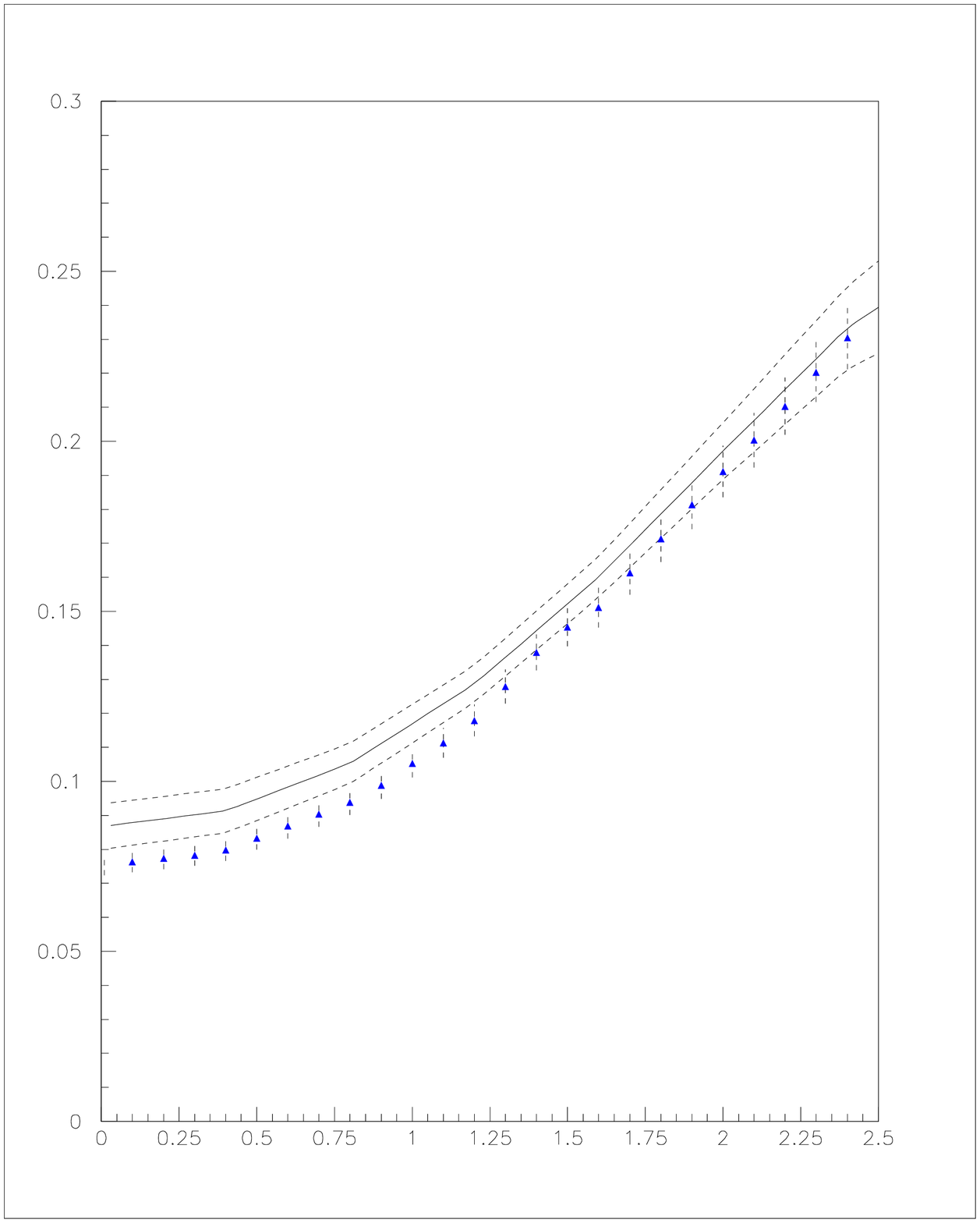,width=0.3\textwidth,height=4cm}
\epsfig{figure=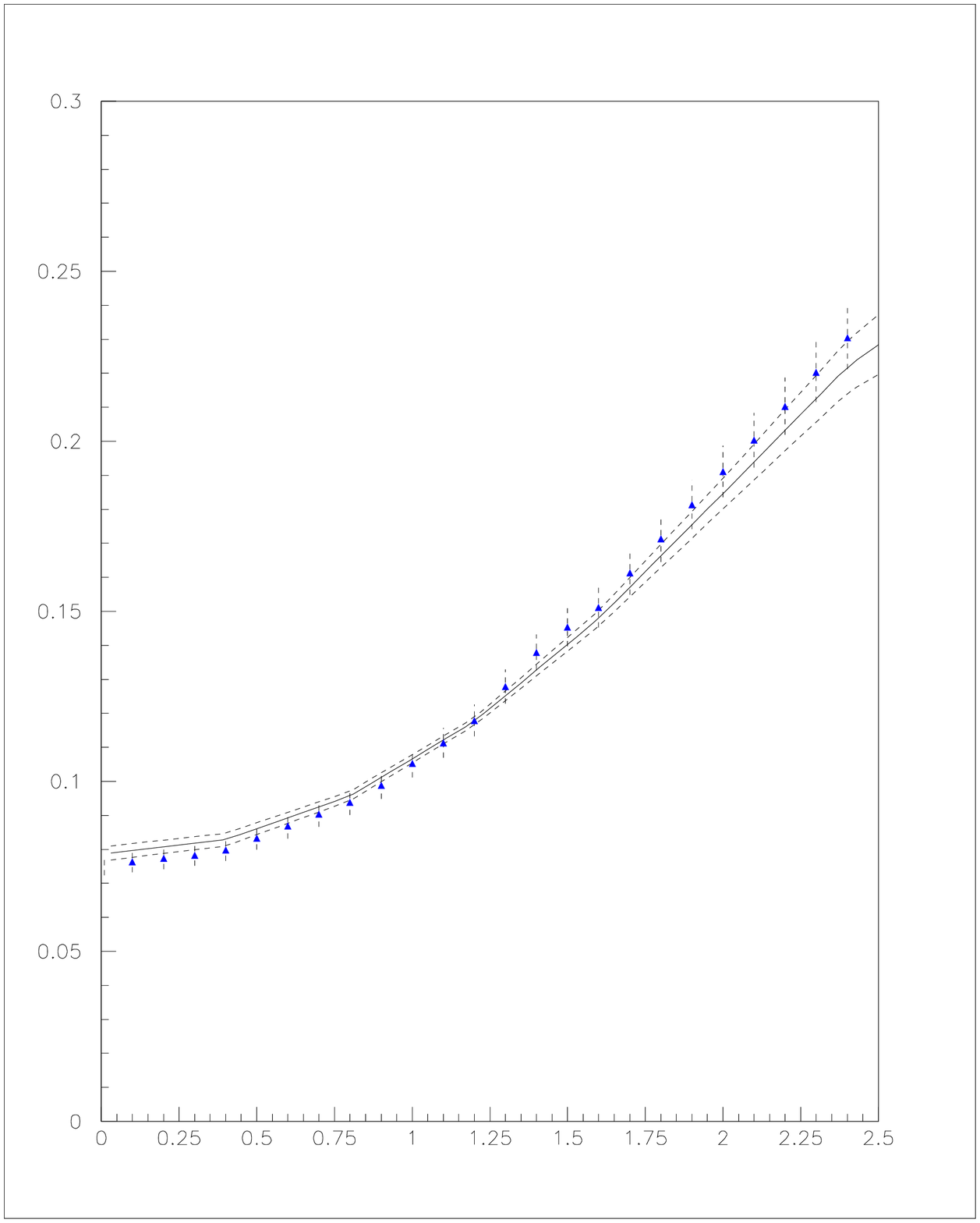,width=0.3\textwidth,height=4cm}
}

\caption {Left: $A_l$ rapidity spectrum generated from MRST04 PDFs,  
compared to the analytic prediction
using ZEUS-S PDFs. Right: the same $A_l$ rapidity spectrum compared to the analytic 
prediction after including these lepton pseudo-data in the ZEUS-S PDF fit.}
\label{fig:asymfit}
\end{figure}

A similar study has been made for the lepton asymmetry. Whereas pseudo-data generated with 
CTEQ6.1 PDFs is in agreement with the ZEUS-S PDFs, pseudo-data generated with MRST04 PDFs 
is not, as illustrated on the left side of Fig.~\ref{fig:asymfit}. The right side of the 
figure shows the result if these MRST04 pseudodata are included in the ZEUS-S PDF fit.
A good fit can only be obtained if the form of the valence parametrizations is relaxed to 
$xf(x)= A x^a (1-x)^b (1+d\surd{x}+c x)$.
The central value of the prediction shifts and the PDF uncertainties are reduced.  
Further information on valence distributions at larger $x$ can be obtained by making 
measurements at high rapidity($y \sim 4$), see \cite{url1,url2}.

\section{Problems with the theoretical predictions at small-$x$?}

However, a caveat is that the current studies have been performed using 
standard PDF sets which are extracted using NLO QCD in the DGLAP~\cite{ap} 
formalism. The extension to NNLO is 
straightforward. However, there may be much larger  
uncertainties in the theoretical calculations because the kinematic region involves  
low-$x$.  There may be a need to account for $ln(1/x)$ resummation (first considered in the 
BFKL~\cite{lip} formalism) 
or high gluon density effects. See reference~\cite{dcs} for a review. 

\begin{figure}[tbp] 
\vspace{-1.0cm}
\centerline{
\epsfig{figure=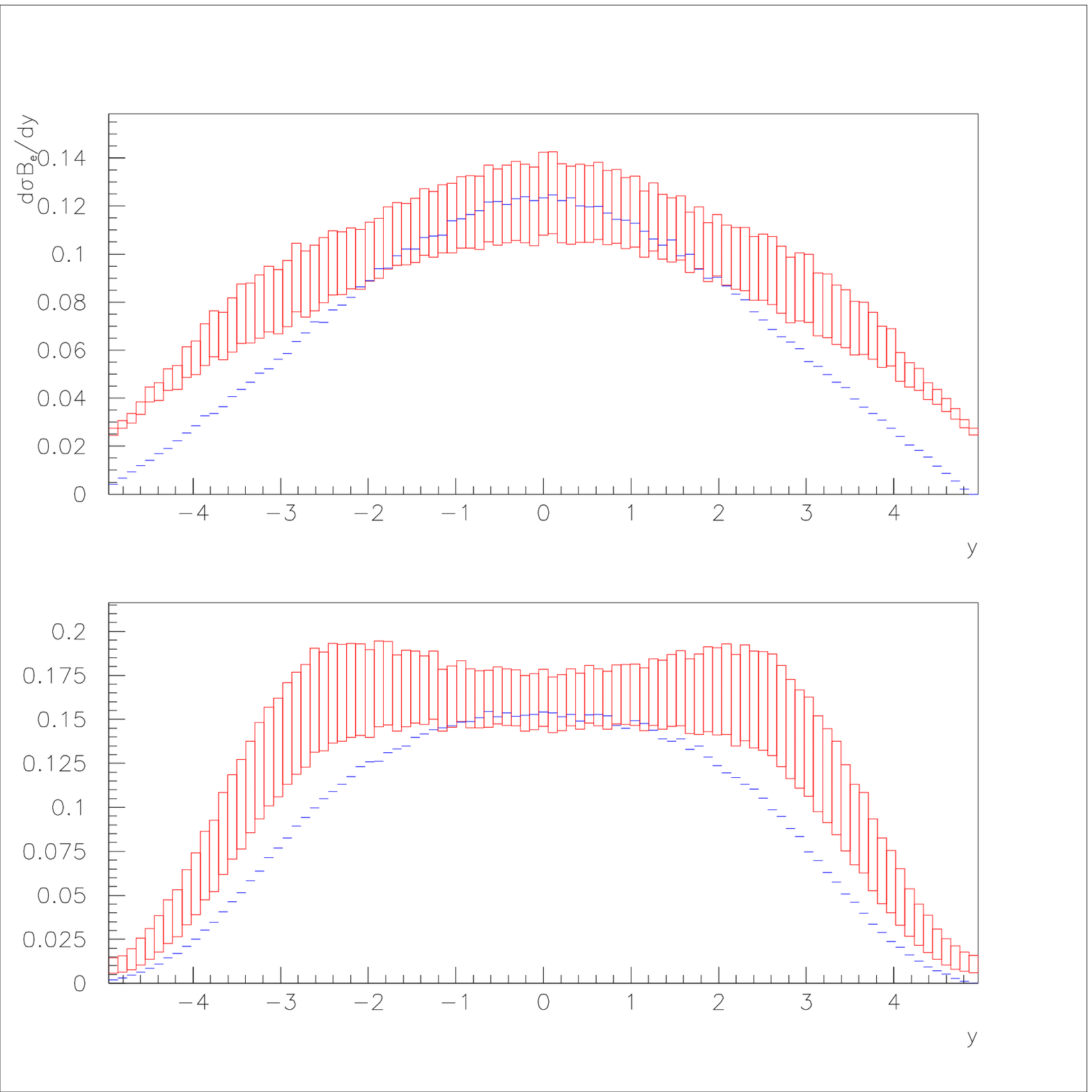,width=0.3\textwidth,height=4cm}
\epsfig{figure=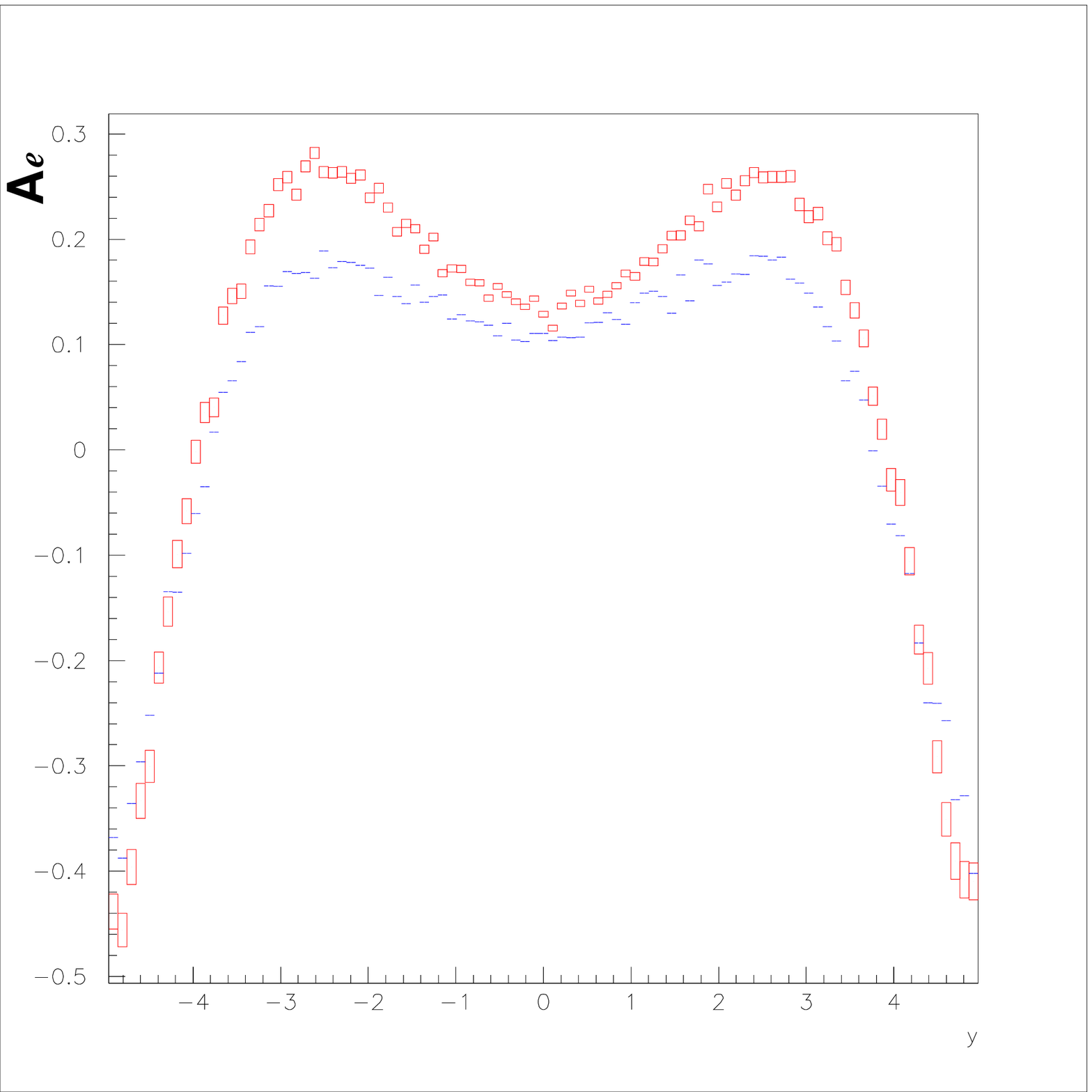,width=0.3\textwidth,height=4cm}
}
\caption {LHC $e^+,e^-,A_e$ rapidity distributions for the MRST03 (blue) compared to CTEQ6.1 
(red) PDFs. (Thanks to A Tricoli)}
\label{fig:mrst03pred}
\end{figure}

The MRST group recently produced a PDF set, MRST03~\cite{mrst03}, which does not include any data for 
$x < 5\times 10^{-3}$. The MRST03 PDF set is thus free from bias due to
the inappropriate use of the DGLAP formalism at small-$x$. 
BUT it is also only valid to use it for $x > 5\times 10^{-3}$. 
What is needed is an alternative theoretical formalism
for smaller $x$. Nevertheless, the MRST03 PDF set may be used as a toy PDF set, 
to investigate the kinematic region where we might  expect to see differences due 
to the need for an alternative formalism at small-$x$. 
Fig.~\ref{fig:mrst03pred} compares predictions using MRST03 and  CTEQ6.1 PDFs. 
The difference in these PDFsets would become evident with data from just 6hours of running.
It might be most fruitful to look for unconventional contributions to low-$x$ physics by 
looking at $W,Z$ $p_t$ spectra rather than at the rapidity spectra. A recent calculation of 
the effect of a lack of $p_t$ ordering on these spectra has shown significant differences 
from the conventional calculations~\cite{resbos}, see~\cite{url1}.

\section{Impact of PDF uncertainties on discovery physics}

\begin{figure}[tbp] 
\epsfig{figure=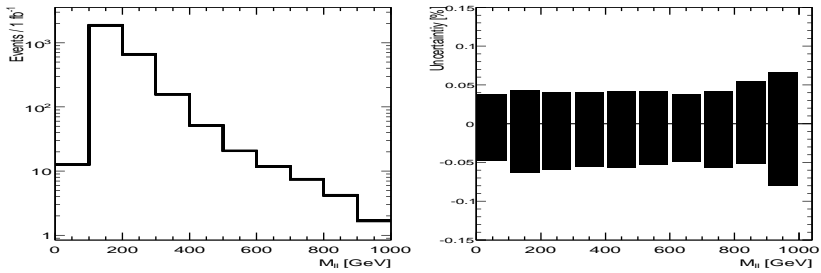,width=0.8\textwidth,height=4cm}
\caption {Left: Di-lepton mass spectrum from 50-1000 GeV Right: 
PDF uncertainties on this spectrum from CTEQ6.1 PDFs. (Thanks to F Heinemann)}
\label{fig:drellyan}
\end{figure}

There are classes of discovery physics which are not much compromised by PDF uncertainties.
For example, the discovery of SM Higgs as discussed in~\cite{ferragh}, see~\cite{url1}. 
There is also recent work on PDF uncertainties on high mass di-lepton production 
which suggests that these uncertainties would not mask $Z'$ production, 
see Fig~\ref{fig:drellyan}. 

The class of BSM physics which would be compromised by PDF uncertainties is anything which 
would appear in the high-$E_T$ jet cross-sections, such as new physics which 
can be written as a contact interaction. Indeed 
in 1996 the high-$E_T$ jet data at the Tevatron~\cite{TeVjets} were first taken as a signal 
for such new physics. 
However, analysis of PDF uncertainties established that these data lie well 
within the current level of PDF uncertainty. The main contribution to the uncertainty on the 
high-$E_T$ jet cross-section comes from the high-$x$ gluon~\cite{cteqjet}. This uncertainty 
also affects high-$E_T$ jets at the scales accessible at the LHC, see for 
example~\cite{cteqjet} and \cite{ferrage} where a study is made of the possibility to 
distinguish compactified extra dimensions. PDF uncertainties reduce the compactification 
scale which can be accessed from $6 TeV$ to $2 TeV$.

\begin{figure}[tbp] 
\vspace{-2.5cm}
\centerline{
\epsfig{figure=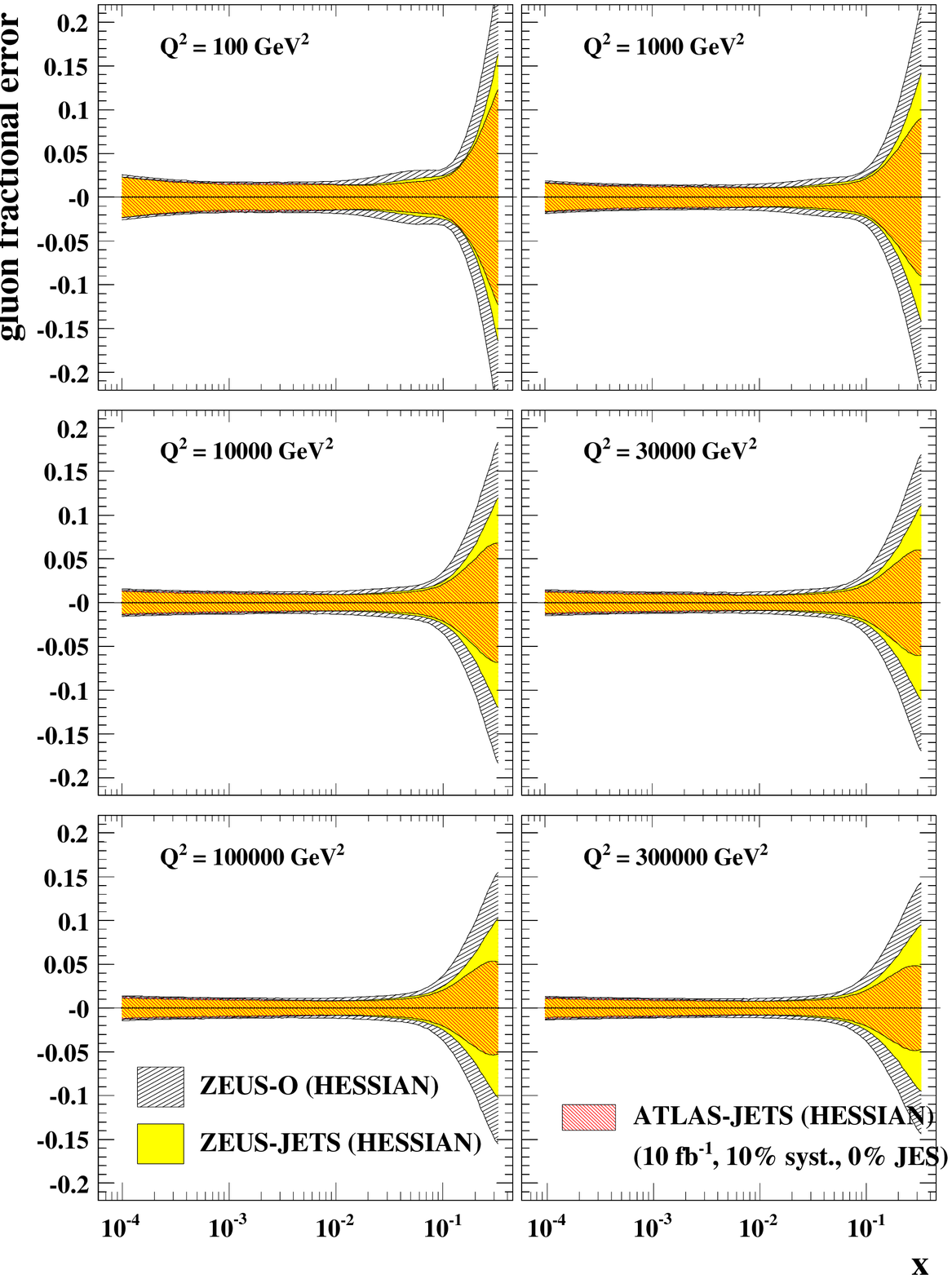,width=0.3\textwidth,height=4cm}
\epsfig{figure=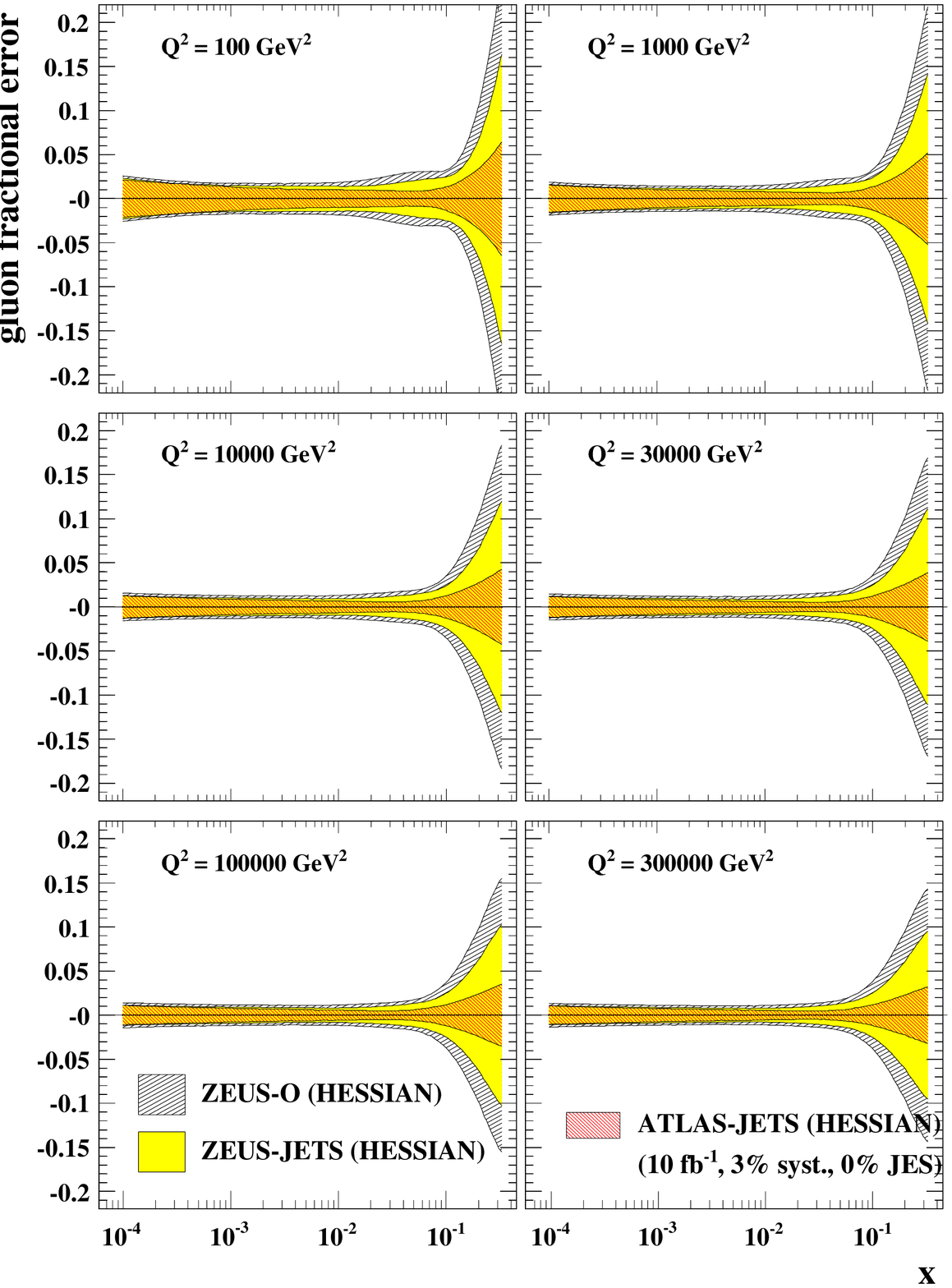,width=0.3\textwidth,height=4cm}
}
\centerline{
\epsfig{figure=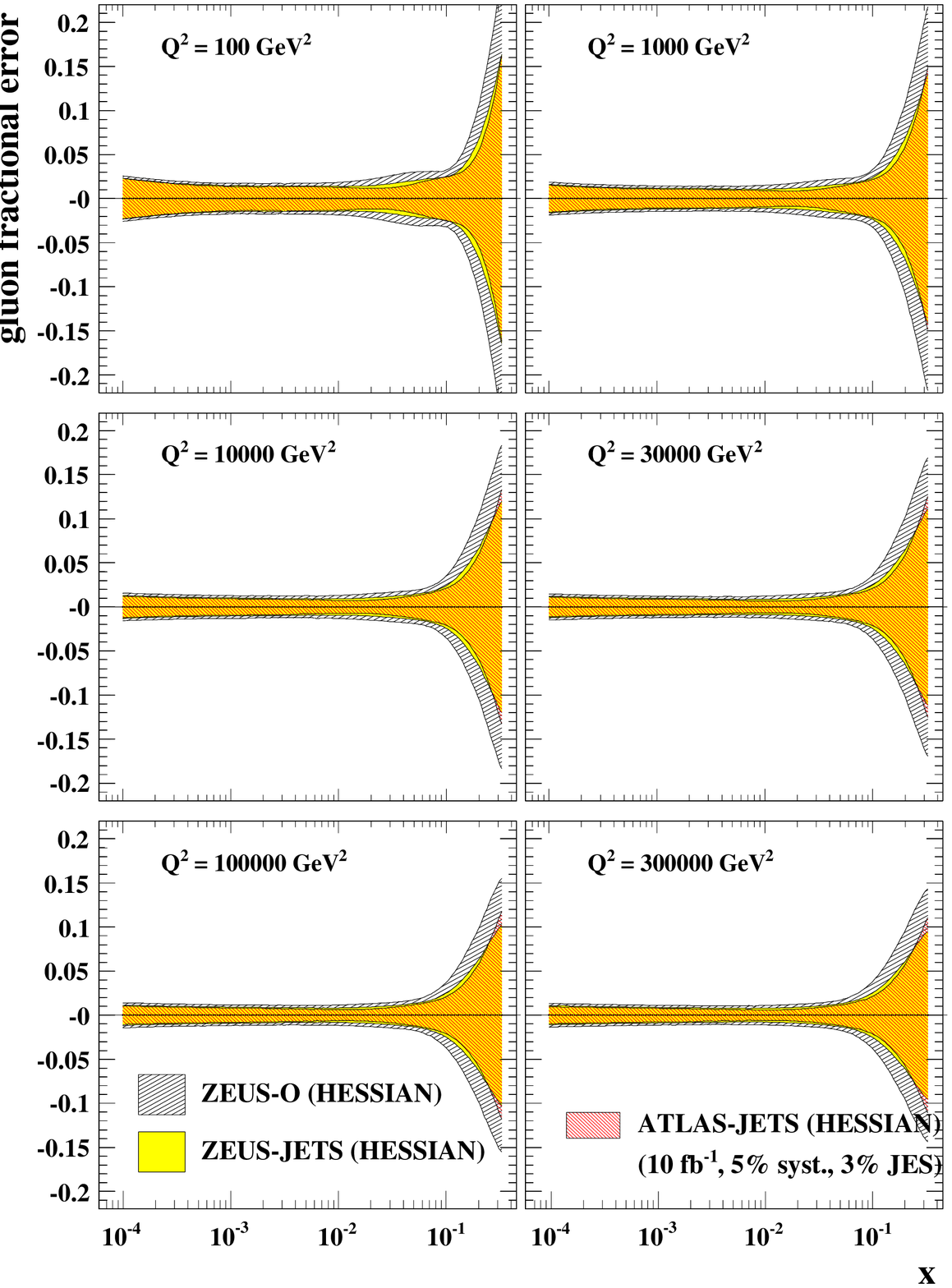,width=0.3\textwidth,height=4cm}
\epsfig{figure=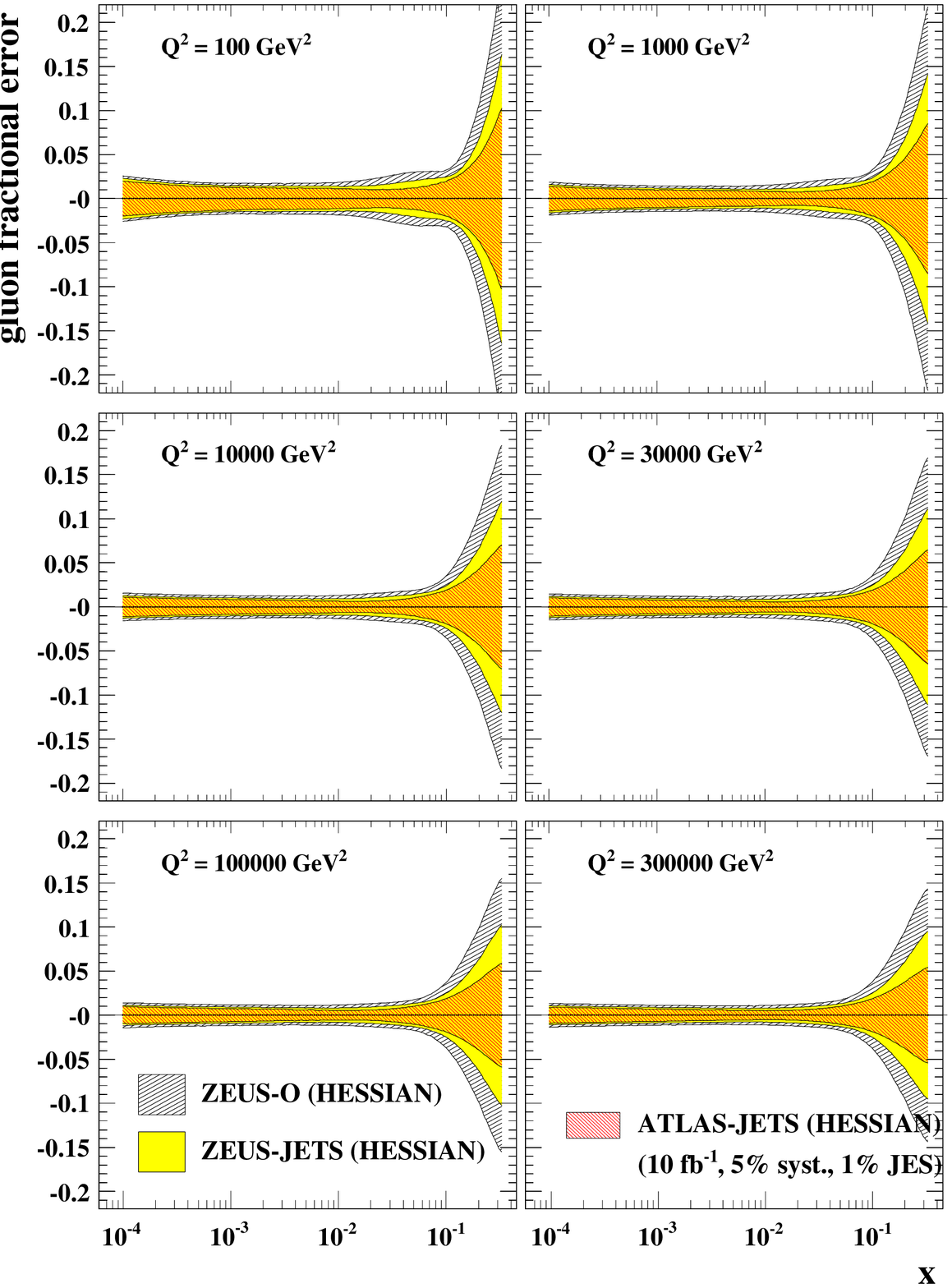,width=0.3\textwidth,height=4cm}
}

\caption {Fractional uncertainty on the gluon PDF at various $Q^2$ for the ZEUS PDF fits.
ZEUS-O uses just ZEUS inclusive cross-section data, ZEUS-JETS adds ZEUS jet production data 
and ATLAS-JETS adds $10 fb^{-1}$ of ATLAS jet pseudodata. Top left: ATLAS pseudo-data with 
$10\%$ uncorrelated systematic errors. Top right: ATLAS pseudo-data with 
$3\%$ uncorrelated systematic errors. Bottom left: ATLAS pseudo-data with 
$5\%$ uncorrelated systematic errors and $3\%$ correlated jet energy scale error.
Bottom left: ATLAS pseudo-data with 
$5\%$ uncorrelated systematic errors and $1\%$ correlated jet energy scale 
error. (Thanks to C Gwenlan)}
\label{fig:jets}
\end{figure}

It is clearly of interest to investigate if LHC jet measurements can 
themselves be used as an input to a PDF fit to decrease this level of uncertainty. Recently 
grid techniques have been developed to input predictions of NLO jet cross-sections to PDF 
fits~\cite{zeujets,fastnlo} and this technique can be used for LHC high-$E_T$ jet 
cross-sections. Data at lower $E_T$ and central rapidity are used, where new physics 
effects are not expected. The programme NLOJET was used to generate grids of parton 
sub-process cross-sections up to $p_T < 3 GeV$, 
in pseudorapidity ranges $0 < \eta < 1$, $1< \eta < 2$, $2 < \eta < 3$. These sub-process 
cross-sections can then be multiplied by the PDFs which are varied in the fit.
LHC high-$E_T$ jet pseudodata was generated using JETRAD and input to the ZEUS-JETS PDF fit 
using this technique. In Fig~\ref{fig:jets} 
the results are shown for a variety of different ATLAS pseudo-data inputs. All samples assume a luminosity of 
$10 fb^{-1}$ however, luminosity is not a restriction provided $L > 1 fb^{-1}$.  
The top left of the figure shows the effect of using ATLAS pseudo-data with an assumed 
uncorrelated systematic error of $10\%$. This is sufficent to give an 
impressive decrease in uncertainty compared to current ZEUS-JETS PDFs. 
The top right of the figure shows the effect of using ATLAS pseudo-data with an assumed 
uncorrelated systematic error of $3\%$. The decrease in uncertainty between these two 
samples is striking. However, on the bottom of the figure the effect of assuming a 
correlated systematic uncertainty 
due to jet energy scale is illustrated. On the left hand side a jet energy scale of $3\%$ is assumed and on the right hand side a jet energy scale of $1\%$ is assumed.
Even a modest $3\%$ jet energy scale is enough to destroy the improvement we had gained. 
The jet energy scale would need to be as small as $1\%$ in order to see any improvement.
In view of this pessimistic conclusion it is of great interest to anticipate future 
improvements in our knowledge of the gluon from analysis of HERA-II inclusive cross-section 
and jet data~\cite{gwenlan}


\begin{footnotesize}

\end{footnotesize}


\end{document}